%% file: 00-main.tex
\PassOptionsToPackage{table}{xcolor}
\documentclass[acmsmall, nonacm]{acmart}

\AtBeginDocument{%
  \providecommand\BibTeX{{%
    \normalfont B\kern-0.5em{\scshape i\kern-0.25em b}\kern-0.8em\TeX}}}

\setcopyright{none}

\usepackage{balance}       
\usepackage{tabularx}
\usepackage{color}
\definecolor{lightgray}{gray}{0.9}
\usepackage{array}

\usepackage{multirow}

\newcommand{\dataset}[1]{\mathcal{D}^{\text{#1}}}

\definecolor{linkColor}{RGB}{6,125,233}

\begin{document}

\title{Quarantined! Examining the Effects of a Community-Wide Moderation Intervention on Reddit}

\author{Eshwar Chandrasekharan}
\authornote{Both authors contributed equally to this research.}
\email{eshwar@illinois.edu}
\affiliation{%
  \institution{University of Illinois at Urbana-Champaign}
  \streetaddress{4226 Siebel Center for Comp Sci}
  \city{Urbana}
  \state{Illinois}
  \country{USA}
  \postcode{61801-2302}
  }
  
\author{Shagun Jhaver}
\authornotemark[1]
\email{shagun.jhaver@rutgers.edu}
\affiliation{%
  \institution{Rutgers University}
  \streetaddress{4 Huntington St}
  \city{New Brunswick}
  \state{New Jersey}
  \country{USA}
  \postcode{08901}}

\author{Amy Bruckman}
\email{asb@cc.gatech.edu}
\affiliation{%
  \institution{Georgia Institute of Technology}
  \streetaddress{85 5th St NW}
  \city{Atlanta}
  \state{Georgia}
  \country{USA}
  \postcode{30308}
}

\author{Eric Gilbert}
\email{eegg@umich.edu}
\affiliation{%
  \institution{University of Michigan}
  \streetaddress{105 S State St}
  \city{Ann Arbor}
  \state{Michigan}
  \country{USA}
  \postcode{48109}
  }


\begin{abstract}
 Should social media platforms override a community's self-policing when it repeatedly break rules? What actions can they consider? In light of this debate, platforms have begun experimenting with softer alternatives to outright bans. We examine one such intervention called quarantining, that impedes direct access to and promotion of controversial communities. Specifically, we present two case studies of what happened when Reddit quarantined the influential communities r/TheRedPill (TRP) and r/The\_Donald (TD). Using over 85M Reddit posts, we apply causal inference methods to examine the quarantine’s effects on TRP and TD. We find that the quarantine made it more difficult to recruit new members: new user influx to TRP and TD decreased by 79.5\% and 58\%, respectively. Despite quarantining, existing users’ misogyny and racism levels remained unaffected. We conclude by reflecting on the effectiveness of this design friction in limiting the influence of toxic communities and discuss broader implications for content moderation.
\end{abstract}



\maketitle

\input{00-contents}


\bibliographystyle{ACM-Reference-Format}
\bibliography{sample,newrefs}

\end{document}

%% file: 00-contents.tex
\input{01-introduction}
\input{02-relatedWork}
\input{03-data}
\input{04-methods}
\input{05-results}
\input{06-discussion}
\input{07-conclusion}

%% file: 01-introduction.tex
\fbox{
\begin{minipage}{36em}
  \textbf{CONTENT WARNING}: To describe the content on TRP and TD, we include examples of misogynistic and racist content that may offend our readers. We include them because these samples help explain Reddit's response and ground our research.
\end{minipage}
}
\section{Introduction}

Over the past few years, many scholars and activists have persistently called for social media platforms to remove content that promotes coded racism, misogyny, and conspiracy theories \cite{suzor2018evaluating}. At the same time, there is a growing recognition that platforms need to better facilitate critical discussions on sensitive social issues like race, masculinity and immigration. 
In light of the hotly contested arguments about when and how they should intervene \cite{Carlson2020,gillespie2015platforms}, platforms have begun to consider the use of softer alternatives to outright bans. In this paper, we examine the influence of one such approach, called \textit{quarantining}, deployed by the popular social media platform Reddit. 

\begin{figure*}
\centering
\begin{tabular}{cc}
(a) Splash page on the interface warns visitors to TRP. & (b) TRP is not shown in search results on Reddit.\\
\includegraphics[width=60mm]{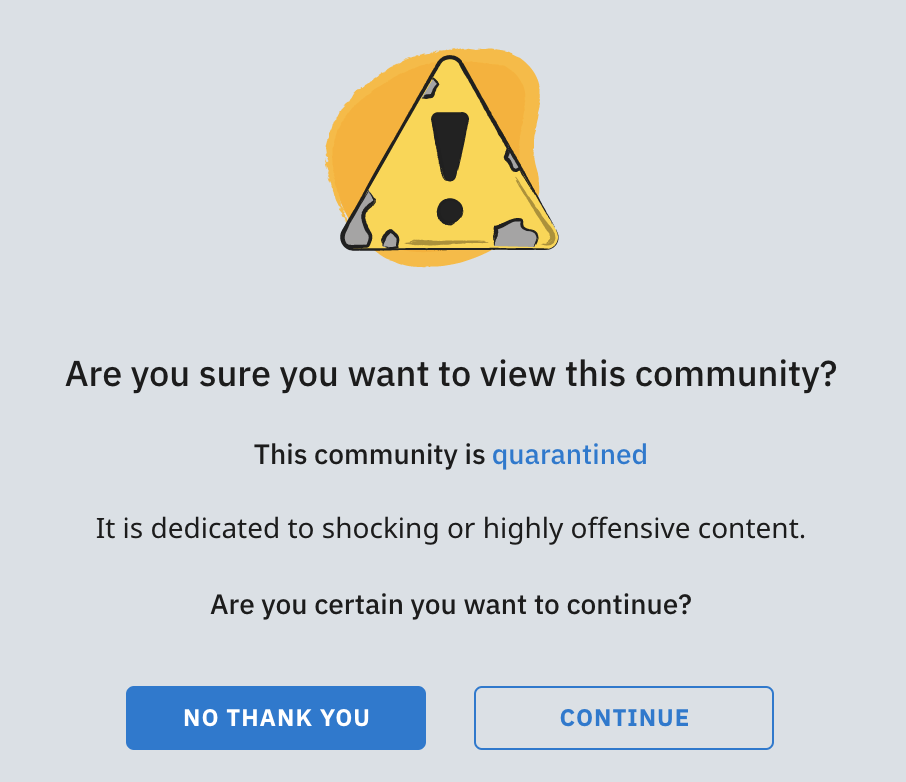} &
  \includegraphics[width=60mm]{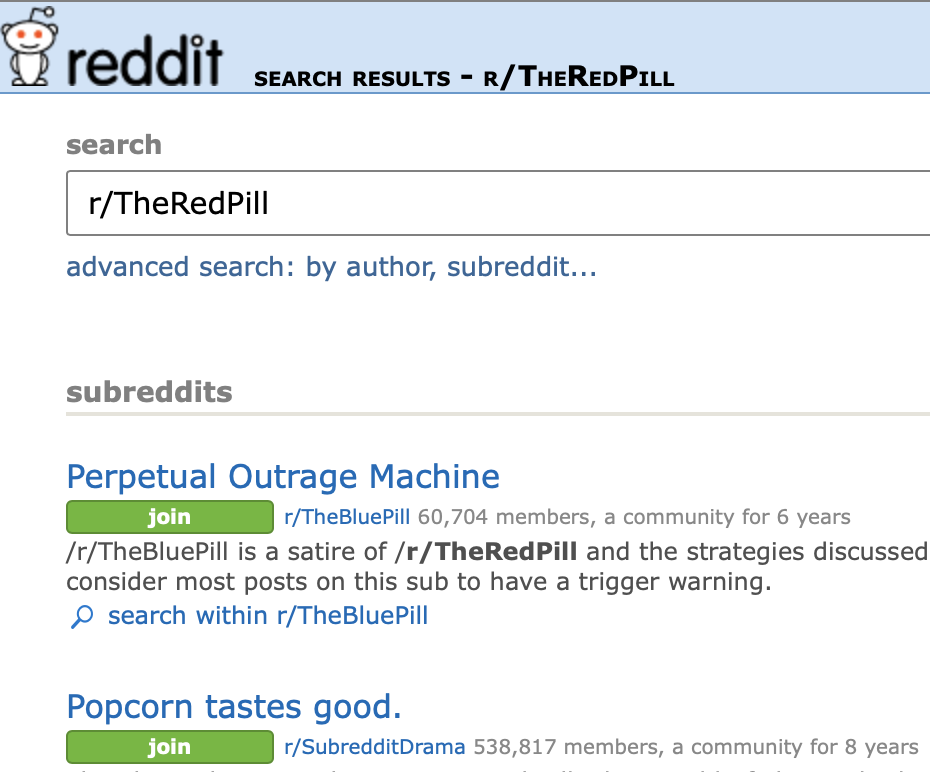} \\ [6pt]
\end{tabular}
\caption{\textit{Design friction} (i.e., intentionally adding points of difficulty to users' interactions with technology) introduced by the quarantine in the form of changes to Reddit's interface, making it harder to find and access TRP content. All users see a new interstitial in Reddit's interface (a), warning them that TRP is quarantined due to its \textit{highly offensive content}. TRP also stopped appearing in Reddit search results during the quarantine (b), and users need to deliberately use TRP's direct URL to access the community. In addition to TRP posts not being indexed by search or the Reddit API, no TRP posts are shown in r/all, a prominent page aggregating the most popular posts from all parts of Reddit. In other words, the quarantine isolated TRP from the rest of Reddit by not surfacing TRP posts to ``unaffected'' users by chance (though users can still access TRP if they so choose).}
\label{fig:quarantine-illustration}
\end{figure*}

Reddit hosts a variety of online communities---thousands of subreddits address topics ranging from books and TV shows to science and politics. Many of these communities are well-moderated, host valuable content, and facilitate constructive discussions, e.g., r/ChangeMyView~\cite{tan2016winning} and r/science~\cite{jones2019r}. Others lack redeeming social value, like the long-since banned r/FatPeopleHate (fat shaming) \cite{chandrasekharan2017you} and r/CreepShots (sexualized images of women without their knowledge). However, other subreddits do not neatly fit into either of these categories. Such subreddits contain content that Reddit considers unsuitable for the wider public but still suitable for members to access as they see fit.
To handle such cases, Reddit deploys the community-wide intervention of \textit{quarantining}. 
When a subreddit is quarantined, visitors trying to access it will see a warning screen, and they must make a deliberate choice to view its content.
Quarantined subreddits also no longer appear in non-subscription-based feeds, recommendations, or search results (see Figure \ref{fig:quarantine-illustration}).

Reddit was explicitly founded on the principles of free speech \cite{stephen_2018}. Therefore, quarantining as a moderation strategy may help the platform fulfill its mission, which is to ``allow just about anything so long as it does not prevent others from enjoying Reddit for what it is: the best place online to have truly authentic conversations''~\cite{stephen_2018}. Quarantining also serves as a warning to the community to monitor itself if it wants to avoid being banned. Softer alternatives to punitive total bans like quarantining are an increasingly deployed option in the toolkit of platform interventions. Considering this, we present in this paper a case-study that examines the effects of quarantining two influential Reddit communities---r/TheRedPill and r/The\_Donald.

\subsection{The r/TheRedPill and r/The\_Donald Quarantine}
Reddit is a content aggregation platform organized into over 1.6 million\footnote{\url{http://redditmetrics.com/history}} user-created communities known as subreddits.
r/TheRedPill (TRP) is one such subreddit, created on October 25, 2012. 
TRP is part of Men's Rights Activists (MRAs), a set of interconnected groups that promote anti-feminist discourse online \cite{nagle2015investigation}.
With over 250,000 subscribers\footnote{\url{https://redditmetrics.com/r/TheRedPill}} at the time of its quarantine,
TRP is currently one of the most influential MRA communities.
Another influential subreddit that was quarantined was r/The\_Donald (TD), a community where users frequently posted offensive content and shared racist and far-right talking points~\cite{koebler_2016, lyons2017ctrl, flores2018mobilizing}. TD was created on June 27, 2015 and had over 750,000 subscribers\footnote{\url{https://redditmetrics.com/r/The_Donald}} at the time of its quarantine.

The following is a typical comment on TRP:
\begin{quote}
\emph{``I believe you are mistaking the message. It's not about the angry 2nd stage TRPers screetching about all women being whores. It's about ignoring these girls' feigned innocence because it's a fucking lie. Better to believe they are whores and not treat them like a fuckin princess. That way they get to earn their place by your side (alongside the other six in your harem).''}
\end{quote}

TD hosts a wide variety of offensive content but is especially known for its racist and Islamophobic posts. A sample TD comment follows:
\begin{quote}
    \emph{``No beards here. Just happy to see liberals triggered. I smile when they cry about gun control and how much they hate Trump. Pass laws just to get them foaming at the mouth. Enjoy niggerfaggot''}
\end{quote}

On September 28, 2018, Reddit announced the quarantine function and on the same day, it \textit{quarantined} TRP; similarly, Reddit decided to quarantine TD on June 26, 2019.\footnote{Note: The authors of this paper are not Reddit employees, and this research was not financially supported or endorsed by Reddit.} 
That is, Reddit introduced \emph{design friction} into its interface when users tried to access either subreddit. The design changes to TRP are shown in 
Figure~\ref{fig:quarantine-illustration}.
When Reddit quarantines a subreddit, visitors to it are shown a splash page that requires them to explicitly opt-in to viewing its content.
Additionally, the quarantined subreddit and its posts stop appearing in Reddit's indexing and search results.
{TD was subsequently banned by Reddit on June 29, 2020. Our study was conducted prior to this shutdown, when the subreddit was still under quarantine and active for participation.
}

In addition to being by far two of the largest subreddits to be quarantined, these communities are also important to study because of the out-sized influence they exerted on Reddit and beyond. r/The\_Donald was the largest community for discussing the candidacy and later the presidency of Donald Trump. Digital outreach, particularly on social media like Reddit, was a particularly important part of the Trump election strategy. On the other hand, online groups like r/TheRedPill play an important role in radicalizing disenfranchised men and producing ideological echo chambers that promote violent rhetoric. Recently, media sources have highlighted the connections between extreme MRAs and mass killings in the U.S. and Canada \cite{hudson_2018,langaas2019have,quinlan_2019}. 
Therefore, it is vital to study how community-wide moderation interventions affect the activities of these communities.

\subsection{Research Questions}
In its official announcements, Reddit states that it quarantines subreddits that ``average redditors may find...highly offensive or upsetting'' or that promote hoaxes or falsifiable conspiracy theories, like Holocaust denial \cite{reddit_quar_2020}.
The goal of quarantining is to prevent users from accidentally viewing  potentially offensive or upsetting content posted on quarantined subreddits ~\cite{landoflobsters2018}. Reddit also expects that quarantining a community may signal the problematic nature of its content to its members, encouraging them to self-monitor their behavior and so escape being banned (i.e., shut down completely). 
We designed our research questions to study how well quarantining achieves these goals.

Quarantining fundamentally differs from banning, which has proven effective in reducing hate-based behavior~\cite{chandrasekharan2017you}.
Banning shuts a space down permanently, forcing its members to leave; prior analyses of community bans have primarily focused on understanding the migration patterns to other online spaces \cite{chandrasekharan2017you}.
On the other hand, the quarantined space, though isolated from the rest of the site, is still available for member participation. 
Quarantining, therefore, censors a community without censoring it \cite{cousineau2020displaced}.
This critical difference sets our work apart from prior work (e.g., [9]). We chose to focus on the immediate effects of quarantine on the ``treated'' subreddits---TRP and TD---given that these forums were still accessible to users.

It is not clear whether the quarantine would ``work'' to achieve the above-stated Reddit goals.
Quarantining, which is essentially labeling a subreddit as ``taboo,'' could lead to a Streisand effect,\footnote{The term ``Streisand Effect'' emerged from an incident where an attempt by the actress Barbra Streisand to restrict online views of her Malibu Mansion on a public website had the paradoxical effect of stimulating public interest and leading to many more views than if she had done nothing
~\cite{jansen2015streisand}.}
i.e., the subreddit could become even more popular due to the quarantine and attract more new users than before.

Reddit's quarantine of TRP and TD provides a rare, natural experiment to examine the long-term effects of isolating radicalized groups.
To examine the effects caused by the quarantine on these two communities, we ask the following research questions:

\begin{itemize}
    \item \textbf{RQ1}: How were the participation levels within TRP and TD affected by the quarantine? 
    \item \textbf{RQ2}: To what extent was the influx of new users to TRP and TD affected by the quarantine? 
    \item \textbf{RQ3}: How was the use of misogynistic and racist language within TRP and TD, respectively, affected by the quarantine? 
\end{itemize}

RQ1 and RQ3 aim to identify the effects of the quarantine on community members participating in TRP and TD (i.e., ``Did Reddit worsen or improve the amount and quality of discourse in the quarantined group?''). RQ2 investigates the effects of quarantining TRP and TD on their ability to attract new users (i.e., ``Did the Reddit quarantine prevent TRP and TD from assimilating new users?'').

\begin{figure*}[!t]
\centering
  \includegraphics[width = 0.75\linewidth]
{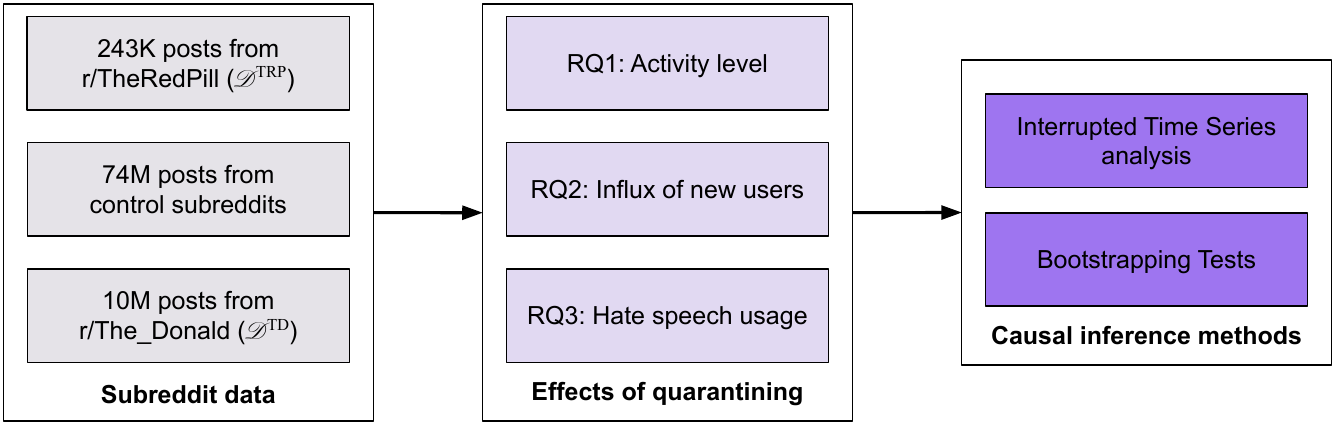}\\ [6pt]
  \caption{Overview of the research pipeline employed in this study. We collected user timelines from the r/TheRedPill (TRP) and r/The\_Donald (TD) subreddits and 100 control subreddits corresponding to each of them for our analyses. Our research questions explored the effects of Reddit's quarantine on posting activity level, influx of new users, and hate speech usage within TRP and TD. We used causal inference methods to answer these research questions.}~\label{fig:research-pipeline}
\end{figure*}

\subsection{Summary of Methods, Findings and Implications}

\emph{Methods.} We answer our research questions by examining observational data from Reddit through a \textit{temporal analysis} of TRP and TD's subreddit timelines. These included all comments and submissions made in the quarantined subreddits six months before to six months after they were quarantined. 
Working with over 85M Reddit posts and comments from Reddit, we chose metrics that include posting volume and frequency of words used from hate speech lexicons.
We then used \textit{causal inference methods} to examine variations in levels of interaction and racism/misogyny, accounting for ongoing temporal trends within TRP and TD in addition to Reddit-wide trends. Figure~\ref{fig:research-pipeline}
 provides an overview of the research pipeline employed in this paper.
 
\emph{Findings.} We determined that the \textit{quarantine severely disrupted the influx of new users} to both subreddits---the rate of new users joining TRP and TD dropped drastically, by over 79.5\% and 58\%, respectively. 
We also found that already assimilated users within both subreddits exhibited changes in posting volume: activity levels of TRP users decreased substantially (by 52.4\% and statistically significant w.r.t. corresponding control subreddits), but activity levels of TD users increased slightly (by 3.8\%, but not statistically significant w.r.t. corresponding control subreddits). 
Despite these changes in posting activity, we found scant changes with respect to levels of misogynistic (TRP) or racist (TD) language. 

\emph{Implications.} For HCI, our work demonstrates the efficacy of introducing simple 
design friction to counteract antisocial behavior in online communities.
For online moderation, we provide a computational framework that internet platforms can adopt to evaluate the effectiveness of moderation interventions.
In her groundbreaking analysis of how to successfully manage common-pool resources, Ostrom argues for using graduated sanctions as a design principle \cite{ostrom1990governing}. 
Our work explores the idea of graduated sanctions in the context of community-wide moderation interventions.
Given the philosophical debates around whether platforms should be allowed to ban certain types of speech \cite{Carlson2020}, quarantining provides a \textit{softer} alternative to banning.
Our findings show one way that platforms can use low-cost design solutions like quarantining to compromise between containing antisocial activities and preserving freedom of speech.

%% file: 02-relatedWork.tex
\section{Background and Related Work}
We next review related work to provide background information about the Manosphere and Red Pill communities, The\_Donald community, online moderation, and platform interventions.

\subsection{The Manosphere}
The \textit{manosphere} is an umbrella term for online communities that hold misogynistic ideas, perpetuate the belief that men are under attack, and propagate anti-feminist discourse \cite{ging2017alphas,hunte2019female,marwick2018drinking}. Over the past few years, the manosphere has gained considerable media attention because of its linkages to the Isla Vista and Oregon mass shootings \cite{chemaly2015mass,dewey2014inside,garkey2016elliot}. 
Campaigns of online harassment, death threats and rape threats against female gamers and journalists have also brought public attention to the manosphere \cite{ging2017alphas,massanari2017gamergate}. 
Prior work has linked these communities to the increasingly influential Alt-Right, a political ideology that embraces misogynistic and white supremacist ideas \cite{kelly2017alt,lyons2017ctrl,mountford2018topic}. 
The manosphere encompasses different communities that focus on a perceived crisis in masculinity, such as involuntary celibates (or incels), men going their own way (MGTOW), pick up artists (PUAs), traditional Christian conservatives (tradCons), and gamer/geek culture communities \cite{schmitz2016masculinities,vito2018masculinity}. 

\subsubsection{The Red Pill communities}
Another salient part of the manosphere are the \textit{Red Pill} communities, which are built on a philosophy that rejects modern feminism and progressivism. 
The Red Pill concept is derived from the 1999 film The Matrix, in which the main character, Neo, is offered a choice to take \textit{the blue pill}, which means living a life of delusion, or \textit{the red pill}, which means becoming aware of the harsh realities of life. 
In the context of the men's rights movement,
the Red Pill philosophy is aimed at awakening men to the purported misandry and brainwashing engendered by feminism \cite{ging2017alphas}. This compelling cultural motif has brought together divergent factions of the \textit{MRA}, enabling a rapid propagation of the Red Pill philosophy across multiple platforms.

The rhetoric of the Red Pill groups is dominated by evolutionary psychology, especially genetic determinism. Proponents use these theories to explain male and female behaviors of sexual selection, often superficially interpreting them to claim that women are inherently hypergamous and irrational and need to be dominated \cite{ging2017alphas}. Red Pill communities prescribe a new form of masculinity that aims to attain the traditional hegemonic aims of sexual conquest and social dominance \cite{mountford2018topic}. They have also `geekified' concepts from evolutionary biology to cultivate a uniquely misogynist, racist, and heterosexist lexicon, including terms like ``beta,''  ``negging,'' ``looksmaxx,'' ``cuck,'' and ``zero night stand'' \cite{ging2017alphas}.

The technological affordances of social media platforms---such as speed, anonymity, disruption of geography, and social disembodiment---have allowed MRA and Red Pill groups to assert male hegemony in new ways \cite{ging2017alphas}. For example, these platforms have facilitated the creation of echo chambers where members unite around emotionally charged claims of victimhood \cite{mountford2018topic}.
Massanari has observed that the algorithmic politics of platforms like Reddit prioritize the interests of young, white, cis-gendered, and heterosexual males and therefore provide fertile ground for anti-feminist activism \cite{massanari2017gamergate}.
The reproducibility and searchability of online content have further contributed to the viral spread of the Red Pill philosophy \cite{johnson1997ethics}.
Since TRP is central to the ecosystem of MRA communities \cite{mountford2018topic}, disrupting its recruitment pipeline would likely have large downstream effects in the manosphere.
In view of this, our work explores whether platforms like Reddit can play an active role in disrupting the spread of misogyny through design frictions like quarantining.

\subsection{The Donald Subreddit}
The\_Donald subreddit (TD) was created on June 27, 2015, around the time that Donald Trump, then primarily known as a business and television personality, announced his presidential run. 
Users on this subreddit regularly discussed Trump-related news, with the vast majority supporting his candidancy and later his presidency.
Ranked as one of the most popular online forums for supporting President Trump \cite{bbc_2019}, TD also heavily influenced the news content on other social media sites like Twitter \cite{zannettou2017web}.

Soon after its creation, this subreddit began causing considerable controversy, both within the Reddit community and in the news media, for its offensive content \cite{flores2018mobilizing}. 
Several Reddit users pointed out the ``hateful'' posts this subreddit hosted and argued that such posts undermined open political deliberation on the platform \cite{flores2018mobilizing}.
TD slowly evolved from being a political community focused on discussing the Trump campaign to becoming a hub for far-right talking points, coded racism, and trolling \cite{koebler_2016}.
Many posters on TD were alt-right users who supported President Trump while frequently participating in racist, sexist, homophobic, and anti-Muslim subreddits \cite{lyons2017ctrl}.
TD members often adopted a ``troll slang'' \cite{Massachs2020} and promoted conspiracy theories, like Pizzagate \cite{ohlheiser_2016} and the Seth Rich murder conspiracy \cite{gilmour_2017}.

On June 26, 2019, Reddit quarantined TD, presumably in light of the problems detailed above. In this work, we examine the after-effects of this quarantine, specifically in terms of change in activity levels and influx of new users. 
Prior research has found that large proportions of users from the influential anti-black, racist subreddit called r/CoonTown migrated to TD after it was banned by Reddit in 2015~\cite{chandrasekharan2017you}.
This migration more than doubled the posting activity of r/CoonTown users within r/The\_Donald~\cite{chandrasekharan2017you}.
Given the influence of subreddits like r/CoonTown on anti-black, racist speech within TD, our study focuses on measuring this toxic impact.

\subsection{Online Moderation and Platform Interventions}
Grimmelmann defines \textit{content moderation} as ``the governance mechanisms that structure participation in a community to facilitate cooperation and prevent abuse'' \cite{grimmelmann2015virtues}. 
Content moderation is a fundamental service that social media platforms provide to shape what content is allowed to remain online and what is removed, the accessibility of different content types, and the actions that accompany content removals \cite{gillespie2018custodians}. It is vital that platforms maintain robust content moderation mechanisms to maintain their reputations and keep existing content usable \cite{jhaver2019automated}.

As platforms grow in size, it becomes increasingly challenging for them to efficiently moderate high volumes of user-generated content efficiently \cite{glaser2018,jhaver2019automated,madrigal2018}. They have therefore established a variety of complex, intricate systems to implement content moderation \cite{taylor2018watch,jhaver2019explanations,jhaver2018blocklists, matias2018civilservant, chandrasekharan2019crossmod}.
In recent years, researchers in HCI as well as other disciplines have started to examine the complexities and challenges involved in these systems \cite{crawford2016,gillespie2018custodians,grimmelmann2015virtues,Lampe2014, fiesler2018reddit, roberts2019behind,seering2019moderator,wohn2019volunteer,chandrasekharan2019hybrid, im2020synthesized}.
Platforms regulate content at different levels of granularity---from moderating individual users or sanctioning an entire community to making site-wide changes---as we review below.

\subsubsection{User-level moderation.}
The bulk of prior research on evaluating content moderation strategies has focused on user-level interventions \cite{asthana2018,kiesler2012,seering:2017}. Kiesler et al. theorized how design principles of clarifying community norms and conferring rewards and sanctions based on user behavior may encourage users to voluntarily comply with norms \cite{kiesler2012}. Seering et al. empirically tested the effectiveness of proactive and reactive moderation tools in preventing antisocial behavior among users \cite{seering:2017}. Several studies have investigated the efficacy of using automated tools to help users make successful contributions on Wikipedia \cite{asthana2018,Morgan2013,warncke2015success,Zhu2018}. More recently, Jhaver et al. evaluated the effectiveness of establishing community guidelines and offering explanations for content removal in improving user attitudes and behaviors \cite{jhaver2019survey,jhaver2019explanations}. 
This extensive literature has helped establish strong theoretical foundations for how user-level interventions can be deployed to improve community outcomes. 

\subsubsection{Community-level interventions.}

While prior research has scrutinized the efficacy of many user-level interventions, we know comparatively little about community-level interventions.
Multi-community platforms can choose to moderate entire communities. For example, Reddit can ban an entire community if it fails to comply with its guidelines\footnote{\url{https://www.redditinc.com/policies/content-policy}}. Alternatively, Reddit administrators can choose to quarantine a community, which forms the focus of the current work. These community-level moderation interventions have widespread implications for a large number of users.
Prior research most closely related to our work studied the effects of banning two controversial Reddit communities and found that banning helped reduce the use of hate speech on Reddit \cite{chandrasekharan2017you}. Some prior work has explored the phenomenon of quarantining on Reddit: Shen and Rose recently conducted a topic analysis of how Reddit users with different political ideologies react to Reddit's quarantine policy \cite{shen2019discourse}. 
Carlson and Cousineau examined the ethics of quarantining using existing ethical frameworks for content moderation and argued that quarantining allows Reddit to continue getting ad revenue from harmful subreddits while appearing to publicly denounce them \cite{Carlson2020}.
We add to this literature by empirically examining a quarantine's effect on activity and hate speech levels of two prominent communities.
We argue that examining the efficacy of quarantining is critical to develop a holistic understanding of how platforms can intervene to disrupt online radicalization even as they allow controversial groups to exist to uphold freedom of speech.

We also contribute to prior literature on graduated sanctions.
In her book on community self-regulation, Ostrom writes that ``graduated punishments ranging from insignificant fines all the way to banishment,
 applied in settings in which the sanctioners know a great deal about the personal circumstances of the other appropriators and the potential harm that could be created by excessive sanctions,
may be far more effective than a major fine imposed on a first offender'' \cite{ostrom1990governing}. Applying Ostrom’s insights to online settings, Kiesler et al. recommend using graduated sanctions against offending users to increase the legitimacy and effectiveness of sanctions in online communities \cite{kiesler2012}.
Our work expands the idea of graduated sanctions to the context of community-wide moderation interventions.

\subsection{Design Friction}
\textit{Design friction} refers to intentionally adding ``points of difficulty encountered during users' interaction with technology'' \cite{Cox2016}. The goal of design friction is to disrupt ``mindless'' automatic interactions and make users reflect before they take action \cite{Cox2016,Mejtoft2019}. Even small barriers prior to an interaction, called \textit{microboundaries}, can help users avoid engaging in behaviors that might not align with their personal values \cite{Cecchinato2015,Cox2016}.

Platforms generally try to facilitate low-friction interactions, e.g., by allowing users to rapidly share photos, access their friends' profiles, or up-vote posts.
However, \textit{some} online friction is highly desirable in certain situations, such as in avoiding the potentially instantaneous damage caused by encountering hate speech \cite{ullmann2020quarantining}. 
Recognizing this, many platforms have deployed moderation interventions that use design friction. For instance, Instagram attempts to moderate the growth of antisocial groups like pro-eating disorder (pro-ED) communities by prefixing search results for  pro-ED hashtags such as \#proana with a content warning \cite{gerrard2018}. 
Search engines display information about suicide helplines as the top results when people query suicide-related information \cite{cheng2019search}.
Pinterest has started blocking search results for some anti-vaccination queries in an effort to combat the anti-vaxxer disinformation \cite{wilson_2019}. 
Reddit itself also uses splash page messages that require users to opt-in to view content on many non-quarantined communities, for example, pornographic (NSFW\footnote{Not Safe For Work}) subreddits. These interventions represent design frictions that constrain access to antisocial content. 

Quarantining on Reddit incorporates two manifestations of design friction. First, it removes the quarantined content from the main /r/all feed and search results, making it accessible only through direct URL link hits. Second, accessing quarantined communities requires an extra click-through acknowledgement, which makes users pause before viewing content in communities they would rather avoid. Both of these design changes introduce friction when users attempt to access the quarantined communities and their content. In other words, the quarantined communities, though still present unlike banned communities, are harder to access.
By examining quarantining, our research aims to engender conversations about the use of design friction to inhibit the growth of offensive speech and radical groups online.

\begin{table}[]
\centering
\caption{Data description. 
$\dataset{}$ denotes the collection of submissions and comments made in r/TheRedPill (TRP) and r/The\_Donald (TD) subreddits in the period from six months before to six months after the quarantine.}
{
\sffamily
\small
\begin{tabular}{@{}lr@{}}
\textbf{Key} & \textbf{Value} \\ \midrule
Number of posts in $\dataset{TRP}$ & 243,758\\
Number of posts in $\dataset{TD}$ & 10,832,057\\ \hline \\[-2ex]
Number of users in $\dataset{TRP}$ & 24,077 \\
Number of users in $\dataset{TD}$ & 173,748 \\ \hline
Number of control subreddits for TRP & 100 \\
Number of control subreddits for TD & 100 \\ \hline
Number of posts from control subreddits for TRP & 19,174,200 \\
Number of posts from control subreddits for TD & 55,414,558
\end{tabular}
}\\ [6pt]
~\label{tab:data-description}
\end{table}

\begin{table*}[!ht]
\centering
\caption{List of top 10 control subreddits for TRP compiled based on the high proportion of co-posting by users from r/TheRedPill. We also provide the description for each subreddit as excerpted from the subreddit page. We compared the effects observed within the TRP subreddit to the effects observed within these subreddits in order to control for Reddit-wide trends. }
{
\small
\sffamily
\rowcolors{2}{gray!25}{white}
\begin{tabular}{r|p{10cm}}
\rowcolor{gray!50}
\hline
\textbf{Control Subreddits} & \textbf{Self-Description}                                                                                                                                                                                \\ \hline
asktrp                      & Here we have Red Pill Discussion for personalized questions about specific situations, people, scenarios.                                                                                           \\ 
askMRP                      & The Red Pill Rangers! Using the Red Pill In Their Marriages!                                                                                                                                        \\ 
The30DayChallenge           & Like all things Red Pill you want to take life by the balls and own it...Be in command of your fitness--that includes your training regimen.                                                          \\ 
marriedredpill              & We are men that subscribe to The Red Pill (TRP) philosophy of sexual strategy, and are dedicated to applying it in marriage or in Long Term Relationships.                                          \\ 
RPChristians                & We are a subreddit that exists to provide a safe place for Christians to discuss problems, solutions and insights on relationships, sex, and biblical masculinity according to biblical principles. \\ 
PurplePillDebate            & PurplePillDebate is a neutral community to discuss sex and gender issues, specifically those pertaining to /r/TheBluePill and /r/TheRedPill.                                                        \\ 
WhereAreAllTheGoodMen       & We're just a bunch of clueless NiceGuys with kindness coins that don't seem to work in women's holes, so that the sex we're entitled to falls out.                                                 \\ 
RedPillWomen                & This community was created as a harbor for RP minded women whose goal is to build a lasting and happy relationship with a great man.                                                                \\ 
DarkEnlightenment           & The Dark Enlightenment, or neoreaction, focuses on the fundamentally flawed tenets of modern western culture.                                                                                       \\ 
BattleOfTheSexes            & A place for no-holds-barred discussion and debate of all things sex and gender.                                                                                                                     \\ 
\end{tabular}\\ [6pt]
}
~\label{table:controlSubsTRP}
\end{table*}

\begin{table*}[!ht]
\centering
\caption{List of top 10 control subreddits for TD compiled based on the high proportion of co-posting by users from r/The\_Donald. We also provide the description for each subreddit as excerpted from the subreddit page. We compared the effects observed within the TD subreddit to the effects observed within these subreddits in order to control for Reddit-wide trends. }
{
\small
\sffamily
\rowcolors{2}{gray!25}{white}
\begin{tabular}{r|p{10cm}}
\rowcolor{gray!50}
\hline
\textbf{Control Subreddits} & \textbf{Self-Description}                                                                                                                                                                                \\ \hline
The\_Europe     & Make Europe Great Again - we update on European right wing leaders and upcoming elections to help the movement.\\
redacted        & [REDACTED]. A wise man once said - {[}removed{]}                               \\
DrainTheSwamp   & Make sure the swamp doesn't win!                                    \\
AskThe\_Donald  & a Pro Trump, Pro Right-wing partisan subreddit                     \\
YouPostOnTheDonald & You post on the The\_Donald you BIGOT REEEE. Links and images of The\_Donald users getting BTFO with the "cuck comeback"                    \\
tucker\_carlson & A subreddit dedicated to discussing Fox news host Tucker Carlson   \\
TheNewRight     & Home of the New Right. The New Right is a big-tent group within New Right Network, dedicated to mobilizing, countering the Left and energizing the Right!                                               \\
kotakuinaction2    & Promoting ethics in journalism, and opposing censorship and political correctness.             \\
HillaryForPrison   & This sub is for everyone who wants to put Crooked Hillary IN FEDERAL PRISON WHERE SHE BELONGS. \\
metacanada      & a place to mock the hive mind of /r/Canada        \\                
\end{tabular}\\ [6pt]
}
~\label{table:controlSubsTD}
\end{table*}

%% file: 03-data.tex
\section{Data}

{We began by reviewing dozens of Reddit forums that discussed quarantining, and TRP and TD were by far the largest and most discussed quarantined subreddits. We therefore chose to focus on these two communities and our analyses only speak to the issues of quarantining these large Reddit communities. Note that Reddit does not provide an official list of all communities that have been quarantined. Therefore, we relied on Reddit search for finding community discussions about quarantining to identify the quarantined communities. We are neither affiliated nor funded by Reddit, and our request to receive a list of all quarantined subreddits to Reddit did not receive any response. 
Though our focus on just two communities is a limitation\footnote{Note that although we present results for just these two communities, our analysis method required us to identify 200 additional control subreddits as well as collect and analyze data from each of them. This is an involved process that included many manual and computationally intensive steps, which made it difficult for us to scale our study to analyze more quarantined subreddits. We chose to conduct a rigorous analysis on two case studies and describe how to scale it to other communities instead of conducting a shallow analysis on a larger number of case studies.}, we hope that our findings and computational approach will inspire further research on alternative approaches to complete bans. We were motivated to take this approach by the success of our prior papers \cite{chandrasekharan2017you,ribeiro2021migration} that examined the effectiveness of banning two subreddits and informed subsequent studies on the efficacy of moderation interventions~\cite{olteanu2018effect, srinivasan2019content, mariconti2019you} and deplatforming~\cite{rogers2020deplatforming,jhaver2020deplatforming}.}

We first collected a dataset of all submissions and comments that were posted on TRP and TD during 2018-2019.
We obtained this data using the Google BigQuery service.\footnote{Example table used for data collection: \url{https://bigquery.cloud.google.com/table/fh-bigquery:reddit\_comments.2018\_03}} In the remainder of this paper, we refer to both submissions and comments as ``posts.''
TRP was quarantined on September 28\textsuperscript{th} 2018 and TD was quarantined on June 26\textsuperscript{th} 2019.
In all further analysis to observe the temporal effects of the quarantine on TRP and TD, we examined 
all the posts made on these subreddits from 180 days before to 180 days after the date of quarantine---April 1\textsuperscript{st} 2018 to March 27\textsuperscript{th} 2019 for TRP and December 28\textsuperscript{th} 2018 to December 23\textsuperscript{th} 2019 for TD.
We denote these posts using $\dataset{TRP}$ and $\dataset{TD}$, respectively.
Table~\ref{tab:data-description} describes our data.

\subsection{Control Subreddits}
The causal inference question in our analysis is whether the quarantining of TRP and TD caused significant changes within them post-quarantine---especially effects on posting activity level, new user influx, and levels of misogyny and racism.
In order to establish causality for any observed effects, we selected subreddits that were \textit{highly likely to be quarantined} as control subreddits in our analyses. 
We use the co-posting behavior of users in our treatment subreddits as a proxy for the likelihood of being quarantined by Reddit.
By comparing the effects observed within TRP and TD to the effects observed within their corresponding control subreddits over the same time period following the quarantine, we obtain lower-bounds for the effects caused within TRP and TD; in other words, this comparison accounts for any Reddit-wide effects observed following the quarantine, thereby establishing the causal link.
Following the approach used in prior work~\cite{chandrasekharan2017you}, we selected our control subreddits as follows:
we compiled a list of all subreddits where treatment users posted pre-quarantine, and then selected the top 100 subreddits based on the percentage of TRP (TD) users posting in these subreddits pre-quarantine (compared to their overall population). We used the PRAW Reddit API\footnote{\url{https://praw.readthedocs.io/en/latest/}} to verify that the top 100 subreddits we selected were still active and had not been banned or quarantined at the time of our analysis.
Tables \ref{table:controlSubsTRP} and \ref{table:controlSubsTD} show the top 10 control subreddits selected through this process. Manual inspection\footnote{The first author used the validation strategy detailed in Sec. \ref{sec:instrument_validation} to annotate whether a random sample of 100 posts from each subreddit contained misogynistic or racist content.} of these subreddits indicated that many of them contained misogynistic and racist content.
Next, for each control subreddit, we retrieved all posts made within the period 180 days before and after the date of quarantine of their corresponding treatment (TRP or TD) subreddit. 
We also retrieved posts submitted within an additional 30 days period prior to the starting date to identify new users in each subreddit.

%% file: 04-methods.tex
\section{Methods}

We divided the timelines of all subreddits into time windows of 10 days each. We assigned a time window index \textit{w\textsubscript{i}} to each time window, with \textit{w\textsubscript{0}} denoting the time window beginning at the date of the subreddit's quarantine (i.e., 09/28/2018 for TRP and 06/26/2019 for TD).
In order to answer our research questions, we focused on three important indicators of user behavior within subreddits: (1) posting activity levels, (2) influx of new users, and (3) usage of toxic speech. Next, we detail how each indicator was operationalized.

\subsection{RQ1: Posting Activity Levels}
We measured posting activity levels within a subreddit using the volume of posts made in each time window \textit{w}. For this, we iterated through each post in the subreddit timeline, and assigned it to the corresponding time window (based on the timestamp indicating when the post was made). 
Next, we aggregated all posts assigned to each time window. The total number of posts contained in each time window represented the subreddit's posting activity level for that time window.

\subsection{RQ2: Influx of New Users}
For each time window \textit{w}, we defined \textit{new users} as those users who had never posted on the subreddit before \textit{w}. To distinguish new from preexisting users, we set aside a buffer period of three time windows (30 days) for which we did not compute new users. We collected all unique users who posted in the first three time windows in a set \textit{seenUsers}. Starting from the fourth time window, we collected unique users in each time window, \textit{currentUsers(w)}, and curated new users in that window, \textit{newUsers(w)}, by compiling those users who were not in \textit{seenUsers} (i.e., \textit{newUsers(w)} is the set difference, \textit{currentUsers(w)} $-$ \textit{seenUsers}). We iterated through each subsequent time window, adding new users seen in that window to the set \textit{seenUsers} (i.e., \textit{seenUsers} = \textit{seenUsers} $\cup$  \textit{newUsers(w)}).
Posting on TRP or TD is a strong indicator of participation, compared to users just ``lurking''~\cite{buntain2014identifying} or viewing content; posts are also more reliable and easier to track.
Note that this analysis operates on a different timescale than our other analyses because of the initial buffer period reserved to identify new users.

\subsection{RQ3: Misogyny and Racism Levels}

\subsubsection{Misogyny lexicon}
Farrell et al. studied the manosphere by investigating the flow of extreme language across seven openly misogynistic subreddits~\cite{farrell2019exploring}. In order to determine the prevalence of misogyny within these communities, they combined theoretical insights from feminist critiques of language to create nine lexicons capturing specific misogynistic rhetoric \cite{farrell2019exploring}.
We used this lexicon consisting of 1,300 misogynistic terms to examine TRP misogyny levels.

\subsubsection{Racism lexicon}
In their study of Reddit bans~\cite{chandrasekharan2017you}, Chandrasekharan et al. created lexicons for detecting racism on the platform by using actual posts made within a racist subreddit banned for harassing behavior---r/CoonTown.
The generation of this lexicon was guided by the European Court of Human Rights' definition of hate speech, including ``comments which are necessarily directed against a person or particular groups of people," focusing on race, religion, and ``aggressive nationalism and ethnocentrism,'' as well as homophobic speech~\cite{weber2009manual}.
Though hate speech and harassment lack specific and all-encompassing definitions, this usage-based approach provides a useful starting point to examine the prevalence of racist speech on different parts of Reddit.
We used this lexicon, which consists of 100 racist terms, to examine racism levels within TD.

\subsubsection{Instrument validation} \label{sec:instrument_validation}
To validate the instrument and ensure that the lexicons actually detected misogynistic and racist content, we manually annotated 100 randomly selected TRP posts and 100 randomly selected TD posts containing at least one term from the misogyny and racism lexicons respectively.
This robustness check helped us answer the following: ``Does the existence of lexicon terms in a post indicate that the overall post is misogynistic or racist in nature?''
One of this paper's authors reviewed each post sampled from TRP and labeled whether the post was misogynistic or not using the definitions adopted by Farrell et al~\cite{farrell2019exploring}.
A similar labeling strategy was used to review posts sampled from TD and determine whether they were racist or not.
Through this validation step, we found that 87 out of the 100 TRP posts containing at least one misogyny lexicon term were indeed misogynistic in nature. In other words, the misogyny lexicon provides a high precision of 87\% when detecting misogynistic posts within TRP.
Likewise, the racism lexicon was found to provide a precision of 82\% when detecting racist posts within TD.
During error analysis, we found that the ``false positives'' were mostly posts that contained only 1-2 lexicon terms and comments that either replied to or quoted information that other authors posted earlier. On average, we found that posts containing greater numbers of lexicon terms were found to be more misogynistic and racist.

\section{Causal Inference Strategy}
The causal inference question we posed was whether the quarantining of a radical community like TRP or TD causes a decrease in radicalization within the subreddit (as measured by posting activity level, influx of new users, and misogyny/racism levels). Ideally, from a study design perspective, Reddit would have randomly chosen the subreddit to quarantine from a list of candidate subreddits. However, TRP and TD were not randomly chosen (to our knowledge), so we employed a number of techniques to approximate the results we would have observed if they had been randomly chosen. To determine whether the post-quarantine effects observed in TRP and TD could be explained by random chance, we calculated $p$-values for many key estimates in this paper. For many of these estimates (namely in RQ1, RQ2), we used classical $p$-values based on asymptotic results.

First, we used a causal inference strategy called \textit{Interrupted Time Series}~\cite{bernal2017interrupted} to measure the causal effects of the quarantine, accounting for ongoing temporal trends within the subreddits. We did this by comparing pre- and post-quarantine differences in posting activity level, influx of new users, and racism/misogyny levels within TRP and TD.
Next, we employed bootstrapping tests to account for Reddit-wide trends. We did this by comparing the effects observed in TRP and TD to those observed in their corresponding control subreddits (examples shown in Table~\ref{table:controlSubsTRP} and Table~\ref{table:controlSubsTD}) that could potentially have been quarantined (see Figures \ref{fig:quarantine-effects-trp} and \ref{fig:control-subreddit-effects-trp} for effects within TRP, and Figures \ref{fig:quarantine-effects-td} and \ref{fig:control-subreddit-effects-td} for effects within TD).

\begin{figure}[!t]
\centering
  \includegraphics[width = 0.5 \linewidth]
{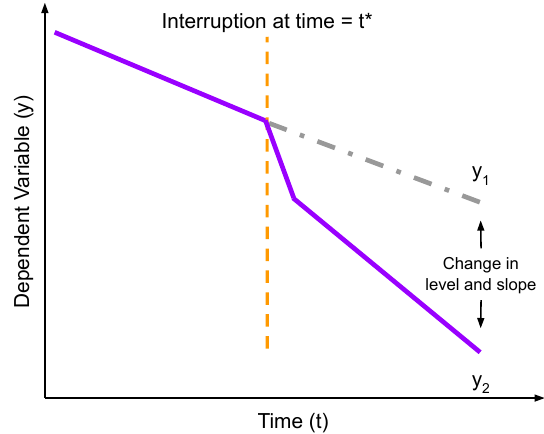}\\ [6pt]
  \caption{Illustration of how the Interrupted Time Series (ITS)~\cite{bernal2017interrupted} strategy accurately estimates changes (in slope and level) caused by the interruption (denoted by the line at $t^*$), by accounting for any ongoing temporal trends within the time series ($y$). In other words, by computing the difference between the slopes and levels of line $y_1$ (extrapolated old series) and $y_2$ (actual new series), ITS analysis estimates the causal effect introduced by the event (interruption) at $t^*$. }~\label{fig:its-explanation}
  \vspace{3pt}
\end{figure}

\subsection{Interrupted Time Series: Accounting for Ongoing Temporal Trends within TRP and TD}
\begin{figure*}[!t]
\begin{tabular}{ccc}
    Activity level (RQ1) & Influx of new users (RQ2) & Misogyny level (RQ3)\\
  \includegraphics[width=45mm]{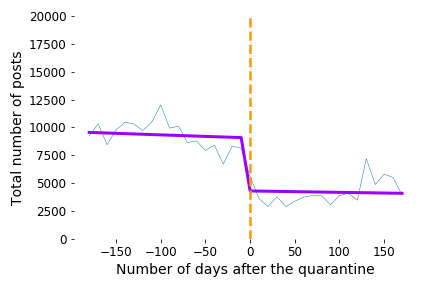} &   \includegraphics[width=45mm]{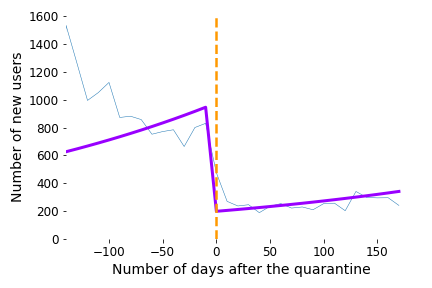} &
  \includegraphics[width=45mm]{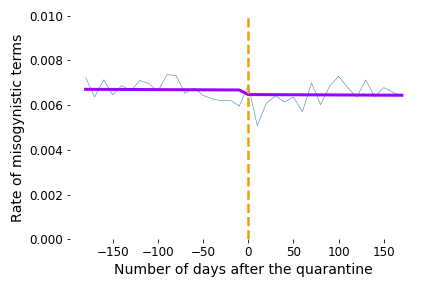}
\end{tabular}\\ [6pt]
\caption{Variations in posting activity levels, influx of new users, and misogyny levels within TRP.
We compute rates for each metric based on all posts from TRP, using time-windows of 10 days, before and after the quarantine.
As shown by the purple lines that ``fit'' the underlying time series, there were dramatic decreases in posting activity level and influx of new users to TRP (both changes were found to be highly significant through ITS analysis and bootstrapping tests), while misogyny levels remained unaffected (we found no significant changes).
{Note that the graphs appear to show a drop even before quarantining because we accumulate the metrics in each 10-day time window at its start, e.g., net posting activity within the first 10 days after quarantining is shown in the graph at time t=0.}}
\label{fig:quarantine-effects-trp}
  \vspace{3pt}
\end{figure*}

\begin{figure*}[!t]
\begin{tabular}{cc}
\centering
    Activity level (RQ1) & Influx of new users (RQ2) \\
  \includegraphics[width=70mm]{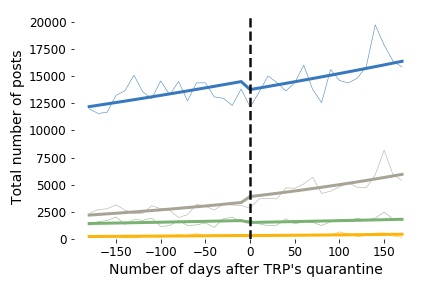} &   \includegraphics[width=70mm]{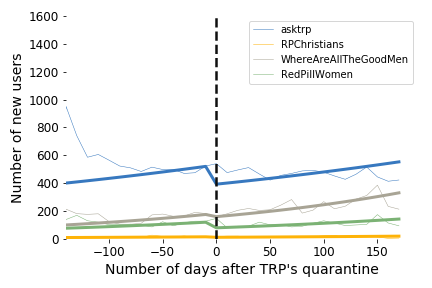} \end{tabular}\\ [6pt]
\caption{Variations in posting activity levels and influx of new users in 4 subreddits selected at random from the top 10 control subreddits for TRP.
We computed rates for each metric based on all posts from 4 random control subreddits, using time-windows of 10 days, before and after the date of TRP's quarantine. We verified that no control subreddit was banned or quarantined. 
Through bootstrapping tests, we determined that the effects observed in TRP were significantly \textit{more extreme} than the effects in control subreddits (visually apparent through comparisons with Figure~\ref{fig:quarantine-effects-trp}).
}
\label{fig:control-subreddit-effects-trp}
\end{figure*}

\begin{figure*}[!t]
\begin{tabular}{ccc}
    Activity level (RQ1) & Influx of new users (RQ2) & Racism level (RQ3)\\
  \includegraphics[width=45mm]{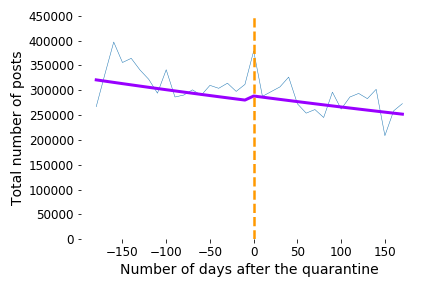} &   \includegraphics[width=45mm]{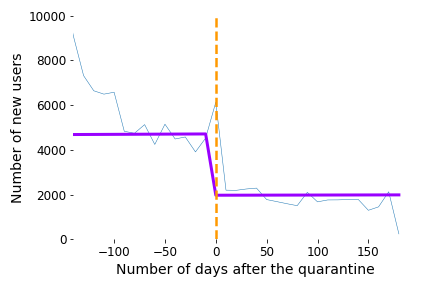} &   \includegraphics[width=45mm]{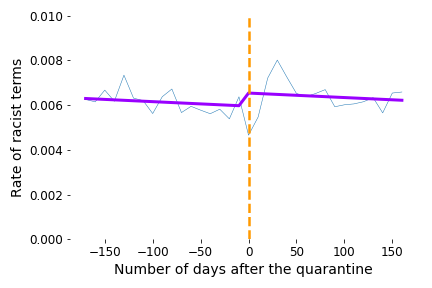}
\end{tabular}\\ [6pt]
\caption{Variations in posting activity levels, influx of new users, and racism levels within TD.
We computed rates for each metric based on all posts from TD, using time-windows of 10 days, before and after the quarantine.
As shown by the purple lines that ``fit'' the underlying time series, there was a dramatic decline in the influx of new users to TD (this change was found to be highly significant using ITS analysis and bootstrapping tests), but there were no drastic changes in posting activity or racism levels.}
\label{fig:quarantine-effects-td}
  \vspace{3pt}
\end{figure*}

The temporal variations in posting activity level, influx of new users, and misogyny/racism levels in TRP and TD are shown in Figure~\ref{fig:quarantine-effects-trp} and Figure~\ref{fig:quarantine-effects-td} respectively.
Based on these figures, we observed substantial drops in posting activity and number of new users in TRP, and a dramatic drop in the number of new users in TD. Note that some drops in activity levels existed even before the quarantine, and these were not caused by the quarantine. There is a possibility that this evidence of a downward trend in activity levels and influx of new users was due to a general trend and not specifically due to the quarantine. To control for this, we performed an Interrupted Time Series (ITS) regression analysis. This let us causally attribute effects in TRP and TD to the quarantine. 

In an ITS analysis, the dependent variable (e.g., posting activity level) for a subreddit is tracked over time, and a regression is used to determine if a \textit{treatment} at a particular point in time caused a change in behavior (i.e., interrupted the time series).
Taking the example of TRP, the \textit{treatment} is the quarantine, initiated on 09/28/2018 (the date the subreddit was quarantined).
As illustrated in Figure~\ref{fig:its-explanation}, an ITS regression models the behavior of the time series before the treatment is applied in order to predict how the series would have looked had the treatment not been applied. This is most often done by regressing the behavior data on time while also including in the regression an indicator variable for the post-treatment time periods (e.g.,~\cite{chandrasekharan2017you}). If we were to simply compare the averages of the behavior pre- and post-treatment without regressing the behavior also on time, then one could easily misinterpret a steadily increasing (or decreasing) time series as a treatment effect~\cite{bernal2017interrupted}. In this sense, an ITS regression shows that the behavior actually changed at the treatment time (e.g., 09/28/2018), instead of being simply a general trend, which could appear to be attributed to any randomly chosen time.

Our analysis focuses on the short-term effects resulting from Reddit's quarantining of TRP and TD. Hence we chose to measure the change in level (as opposed to slope/trends) as our primary metric to examine variations in our ITS analyses. We build our regression models in Python using the statsmodels\footnote{\url{https://www.statsmodels.org/stable/index.html}} Python module. We used poisson regression models for analyzing changes in posting volumes and new user influx, and OLS (Ordinary Least Square) models for analyzing changes in hate speech usage.

\subsection{Bootstrapping Tests: Accounting for Reddit-wide Trends}
We used significance tests to determine whether the observed effects in TRP could reasonably be ascribed to the randomness introduced in selecting the sample. If not, we have evidence that the effect observed in the sample reflects an effect that is present in the population~\cite{hesterberg2005bootstrap, kay2016researcher}. 
In order to account for Reddit-wide trends leading to decreases in activity levels across all subreddits, we developed a $p$-value using a bootstrapping test~\cite{hesterberg2005bootstrap}. 
To conduct this test, we estimated the change in the dependent variables (i.e., posting activity levels, influx of new users) with respect to the date of quarantine for all control subreddits. 

\begin{figure*}[!t]
\begin{tabular}{cc}
\centering
    Activity level (RQ1) & Influx of new users (RQ2) \\
  \includegraphics[width=70mm]{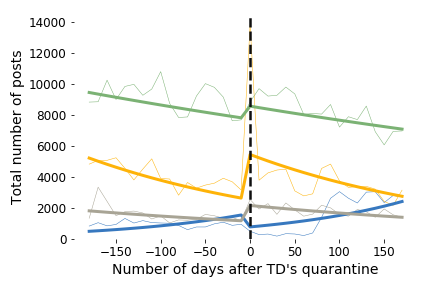} &   \includegraphics[width=70mm]{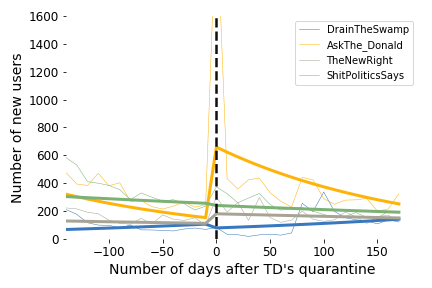}
\end{tabular}\\ [6pt]
\caption{Variations in posting activity levels and influx of new users in 4 subreddits selected at random from the top 10 control subreddits for TD.
We computed rates for each metric based on all posts from 4 random control subreddits, using time-windows of 10 days, before and after the date of TD's quarantine. We verified that no control subreddit was banned or quarantined. 
Using bootstrapping tests, we determined that the reduction in the influx of new users observed in TD was significantly \textit{more extreme} than the corresponding effects in control subreddits (visually apparent through comparisons with Figure~\ref{fig:quarantine-effects-td}).
}
\label{fig:control-subreddit-effects-td}
\end{figure*}

First, we ran the ITS analysis for all 100 control subreddits and obtained a new value of the coefficient for the ITS, $\beta$, for each dependent variable. This step yielded a distribution of 100 coefficients for the ITS coefficient, $\beta$. 
Figures~\ref{fig:control-subreddit-effects-trp} and \ref{fig:control-subreddit-effects-td} illustrate the effects observed in some control subreddits.
Next, these distributions are compared to the actual $\beta_{TRP}$ and $\beta_{TD}$ from our analysis. If the actual $\beta_{TRP}$ was ``extreme'' in relation to the distribution, then this provided evidence that the effect estimate we observed was unlikely to have been produced by chance.
To formalize the notion of ``extreme,'' we used $p$-value, which denotes the proportion of the $\beta$ distribution that exceeds $\beta_{TRP}$. For our analysis, because we were interested in drops in posting activity levels and numbers of new users, $p$ was the proportion of the $\beta$ distribution that was smaller than the actual $\beta_{TRP}$. We bootstrapped by running multiple simulations to compute the distribution of $\beta$s for control subreddits. In each simulation, we randomly sampled 50 coefficients, and computed \textit{how extreme $\beta_{TRP}$ was} when compared to the sampled distribution.
Because the distribution was constructed by simulation, we set up the simulations to terminate when the \textit{p}-value of sampled distributions reached saturation (i.e., no change in $p$-value following consecutive simulations).
We repeated a similar procedure to compute \textit{how extreme $\beta_{TD}$ was} when compared to the effects in its corresponding control subreddits.

\begin{table}[]
\centering
\caption{Interrupted Time Series regression results for posting activity level, influx of new users and the usage of misogynistic/racist language in TRP and TD. The $\beta$ coefficient from ITS regression, as well as the one-tail $p$-values are also included. Results show a highly significant causal effect of the quarantine on the substantial decrease in posting activity levels and influx of new users in TRP. However, there was no significant change in the use of misogynistic language caused by the quarantine. In TD, the influx of new users caused by the quarantine declined appreciably, but there were no drastic changes in posting activity levels or the use of racist language.
Overall, our findings indicate that the quarantined disrupted the incoming streams of new users to both TD and TRP, but it did not change behaviors of preexisting members of these communities. 
Here, p<0.001: ***}
{
\small
\sffamily
\begin{tabular}{@{}l|rr|rr@{}}
\multicolumn{1}{l}{\textbf{Metric}} & \multicolumn{1}{c}{\textbf{$\beta_{TRP}$}} & \multicolumn{1}{r}{\textbf{\% change TRP}} & \multicolumn{1}{c}{\textbf{$\beta_{TD}$}} & \multicolumn{1}{c}{\textbf{\% change TD}} \\ \midrule
Activity level & -0.7418*** & -52.37\% & 0.0367*** & 3.74\% \\
Influx of new users & -1.58*** & -79.51\% & -0.8732*** & -58.24\% \\
Misogyny level & -0.0002 & -0.0 & -0.0000 & -0.0\%
\end{tabular}\\ [6pt]
}
~\label{table:ITSanalysis}
\end{table}

%% file: 05-results.tex
\section{Results}
We now summarize results from our analyses.

\subsection{Changes in Posting Activity Level (RQ1)}

Results from the ITS analysis are shown in Table~\ref{table:ITSanalysis}. 
The `\% change' columns in this table denote the percentage reduction in measured metrics from their estimated values post-quarantine.
We observed that quarantining significantly reduced---by over 52.37\% ($\beta_{TRP}$ = -0.7418, \textit{p}-val < 0.0001)---posting activity in TRP (shown in Figure~\ref{fig:quarantine-effects-trp}). By comparing change in posting activity across control subreddits using bootstrapping tests, we found that a one-sided significance test gave \textit{p}-val < 0.0001, indicating strong evidence that the TRP activity decline was indeed caused by the quarantine, not by a general Reddit-wide decrease in posting activity.
On the other hand, we observed that quarantining significantly increased---by over 3.74\% ($\beta_{TD}$ = 0.0367, \textit{p}-val < 0.0001)---posting activity in TD (shown in Figure~\ref{fig:quarantine-effects-td}).
When we compared change in activity level across control subreddits using bootstrapping tests, we found that a one-sided significance test gave \textit{p}-val > 0.05. This indicates that the increased posting activity in TD was not extreme compared to control subreddits. Our manual analysis of post-quarantining posts indicates that this slight increase may be due to the new discussions about the subreddit being quarantined.

\subsection{Sharp Decreases in the Influx of New Users (RQ2)}

Through the ITS analysis (results in Table~\ref{table:ITSanalysis}), we observed that the quarantine caused a significant decrease of over 79.51\% ($\beta_{TRP}$ = -1.58, \textit{p}-val < 0.0001) in the influx of new users to TRP and a decrease of over 58.24\% ($\beta_{TD}$ = -0.8732, \textit{p}-val < 0.0001) in the influx of new users to TD (shown in Figure~\ref{fig:quarantine-effects-trp} and Figure~\ref{fig:quarantine-effects-td}, respectively). 
Comparing the change in rate of new users posting within different subreddits using bootstrapping tests, we found that a one-sided significance test gave \textit{p}-value < 0.0001 for both TRP and TD. This provides strong evidence that the decreases in influx of new users to TRP and TD were indeed caused by the quarantine, not by a general Reddit-wide decrease in new user influx.
These results suggest that the quarantining of TRP and TD did not cause users from other factions of Reddit to post on TRP and TD, indicating an absence of the ``Streisand effect.'' Thus, in the cases of r/TheRedPill and r/The\_Donald, Reddit's intervention of quarantining drastically reduced the number of new, unaffected users posting on TRP and TD.

\subsection{No Significant Effects on Hate Speech Usage (RQ3)}
Through the ITS analysis, we found that the quarantine did not cause a significant change in the usage of misogynistic words in TRP ($\beta_{TRP} = -0.0002$, \textit{p}-val > 0.5) or racist words in TD ($\beta_{TD} = -0.0000$, \textit{p}-val > 0.5).
When using all 1,300 words in the misogyny lexicon, we observed a slight downtick in the use of misogynistic terms immediately after the quarantine (shown in Figure~\ref{fig:quarantine-effects-trp}). Upon manual inspection, we determined that most posts following the quarantine focused on the event itself (i.e., Reddit's decision to quarantine TRP), criticising the admins for their actions, and planning their next steps in case Reddit decided to ban TRP (i.e., shut TRP down permanently, similar to bans imposed on hateful subreddits in the past~\cite{chandrasekharan2017you}). This caused the slight decline in the use of misogynistic words immediately after the quarantine. However, this decrease in misogynistic speech was not significant, and it was only temporary. Over time, TRP users resumed their misogynistic speech, leading to levels of misogyny that were comparable to the pre-quarantine time period. In other words, the level of misogynistic language in the TRP community was not affected by the quarantine. 
Likewise, the levels of racism prevalent on TD was not affected by the quarantine.
Again, we observed a slight uptick in the use of racist words in the TD subreddit immediately after the quarantine, but this increase was not significant, and it was only temporary. 
Since we found no significant trends in our ITS analyses for hate speech usage in TRP and TD, we did not conduct the follow-up bootstrapping analyses.

\subsection{Additional Checks}
{We verified that quarantining TRP and TD did not create conditions within control subreddits that skewed the baseline coefficient distribution in the opposite direction. 
For example, the \textit{extremeness} (i.e., \textit{p}-val and \% reduction) of the decrease in the rate of new users joining TRP, as computed through the bootstrapping tests, was not caused by increases in new user influx within all of the control subreddits.
In order to check for this, we ran the ITS analysis for posting volume and new user influx within all control subreddits (100 for TRP and 100 for TD).
We found that only 9 out of the TD's 100 control subreddits had a significant increase in posting volume (wrt the date of TD's quarantine), while 15 out of TRP's 100 control subreddits had a significant increase in posting volume.
Additionally, only 2 out of the TD's 100 control subreddits had a significant increase in new user influx while none of TRP's 100 control subreddits had a significant increase in new user influx.
Therefore, the \textit{extremeness} of the observed decreases in posting volume within TRP and new user influx to TRP and TD were not inflated due to a simultaneous increase in these variables within all of the control subreddits.}

%% file: 06-discussion.tex
\section{Discussion}

We now summarize what our findings reveal about the effectiveness of quarantining as a moderation strategy to curb the flourishing of hateful groups online. We also highlight the theoretical and design implications of this research, discuss its limitations, and identify directions for future work.

\subsection{The Effectiveness of TRP's and TD's Quarantine} \label{sec:discussionEffectiveness}

\subsubsection{Posting activity in TRP dropped drastically, but remained misogynistic. Posting activity in TD slightly increased, and remained racist.} Through our ITS analysis, we found that the quarantine significantly reduced posting activity in TRP.
Despite this reduction, the relative use of misogynistic terms in TRP posts showed no significant change.
Additionally, quarantine slightly increased the posting activity in TD and the relative use of racism terms in its posts showed no significant change.
This lack of change in the use of toxic terms suggests that users who are already entrenched within radical groups resist changing their behavior. As mentioned earlier, Reddit noted that one of the primary goals of quarantining is to compel users to rethink their behavior and reduce offensive posts. 
For example, when quarantining TRP, Reddit administrators messaged its moderators that to exit their quarantine status, the community would need to demonstrate a sustained reduction of misogyny.
However, our findings suggest that quarantining did not serve this goal.
A lack of reduction in posting activity in TD post quarantine also indicates that quarantining a community does not consistently deter its activity levels. Thus, if the platform's goal is to curb activity in an offensive community, quarantining by itself is not a reliable solution.

\subsubsection{Influx of new users to TRP and TD declined.} Before we began our analyses, it was not clear how quarantining would affect the stream of new users into quarantined sites. Labeling a subreddit as ``highly offensive'' could have resulted in a Streisand effect~\cite{jansen2015streisand}, i.e., making it even more popular and therefore attracting more new users than before. However, we found that this was not the case: despite widespread discussions about the quarantine in multiple Reddit communities, the numbers of new users to TRP and TD significantly dropped due to the quarantine. 
This demonstrates how design friction---indicating to new users the offensiveness of a group and introducing barriers to the access of offensive spaces---can substantially disrupt the growth of such groups.

In recent years, internet platforms have often been accused of serving as a venue to radicalize disenfranchised groups \cite{archetti2013terrorism,neumann2013options}. 
However, the effectiveness of quarantining as a strategy to impede the integration of new individuals into TRP and TD communities offers important empirical evidence of how platform interventions can successfully address these societal problems.
Platforms like Reddit have thousands of active user-created communities, which makes monitoring the activity of every community challenging. Nevertheless, our findings suggest that platforms can provide societal value by devoting resources to detecting, labeling and isolating antisocial groups.

\subsubsection{Cross-subreddit effects of quarantining.} 
Through the ITS analysis within the control subreddits, we found there were no drastic increases in posting volume and new user influx within these subreddits that were caused by the quarantine.
In fact, there was a decrease (on average) of 40.78\% in posting volume and a decrease of 41.47\% in rate of new users posting within TD's control subreddits. 
Similarly, there was a decrease (on average) of 20.16\% in posting volume and a decrease of 31.69\% in rate of new users posting within TD's control subreddits. 
Moreover, we found that 26 out of the TD's 100 control subreddits had a significant decrease in posting volume (wrt to the date of TD's quarantine), while 60 out TRP's 100 control subreddits had a significant decrease in posting volume.
More importantly, 63 out of the TD's 100 control subreddits had a significant decrease in new user influx while 89 of TRP's 100 control subreddits had a significant decrease in new user influx.
These results provide early evidence that quarantining likely depressed the activity levels and new user influx in many other communities frequented by existing users of TRP and TD.
At the very least, quarantining TRP and TD did not \textit{spread the infection} to other parts of Reddit.

\subsubsection{Calls to migrate elsewhere.} In additional analyses, we explored whether the reduction in activity on TRP and TD could be explained by community managers asking users to migrate on a different site. Moderators usually communicate such official announcements through posts that are stickied at the top of the subreddit. We manually reviewed all posts submitted on TRP and TD between the time these subreddits were quarantined and the end date for our analyses (i.e., six months after the quarantine). We could not find any official (stickied) posts asking users to migrate from TRP within the time period of our study. Indeed, as of Sept 1, 2020, the TRP subreddit still remained accessible. However, we found one post on TD, submitted by a moderator on Nov 19, 2019, that asked users to bookmark an external site \url{https://thedonald.win}, presumably to prepare users to migrate there in case of a ban. To evaluate whether our results are affected by this official announcement on TD, we ran our ITS analysis only up until the time window before this post was submitted. Our results show that even for this restricted period, quarantining significantly reduced the new user influx ($\beta$=-.927, p<=.001). This suggests that even before the official calls for migration to a different site, quarantining had disrupted the activity on TD.

\input{06-implications}

\subsection{Limitations and Future Work}

\subsubsection{Focus on effects within the quarantined communities.} In this study, we examined whether the quarantined subreddit could become more popular due to the quarantine and attract more new users than it had previously. Since the communities still exist, we chose to focus our research questions on understanding the changes within them post-quarantine instead of its members' migration to other communities.
Our results clearly demonstrate the effectiveness of quarantining in delimiting the influx of new users to such problematic forums. Even though these quarantining events were widely discussed on Reddit, we did not observe a Streisand effect of significantly increased user influx.
However, it is possible that quarantining may just shift problematic behavior and lead to increased activity elsewhere. We provide evidence that quarantining did not lead to drastic increases in activity on similar other active subreddits (Sec. \ref{sec:discussionEffectiveness}). Yet, a more rigorous analysis of members' migration to other websites is beyond the scope of this paper and is covered in a different paper \cite{ribeiro2021migration}. In summary, our cross-platform analyses show that quarantining did not significantly increase the posting activity of these communities on other websites.
Going forward, comparative analyses of offensive groups that continue to operate while quarantined on mainstream platforms versus groups that self-host on external websites would substantially inform the use of community-level sanctions.

\subsubsection{Focus on two specific communities.} We focused on two specific quarantined communities, TRP and TD subreddits. To really understand whether quarantining ``works,'' we need to study more cases of quarantined subreddits. Nevertheless, this research examines important aspects of the quarantining of most prominent subreddits, and is therefore a key first step towards answering that larger question.
Amidst the growing amounts of research on online moderation and new ways to counter antisocial behavior \cite{chandrasekharan2019crossmod,cheng2015antisocial,jhaver2018blocklists}, we provide a powerful tool for causally evaluating moderation interventions. 
Our open-source code would also help community managers to evaluate the effects of platform sanctions on their own communities.

A wide range of subreddits get quarantined. Quarantined subreddits include communities that focus on racist tropes, pro-eating disorders, cannibalism, and holocaust denial. We observed that while a few subreddits have their quarantine status removed and return to normal operations after some time, the majority of subreddits are subsequently banned. 
It would be interesting to analyze whether the subreddits that come out of quarantine maintain improved discourse over long periods of time or whether they return to incendiary behavior.

\subsubsection{Lexicons for measuring hate speech.} We applied several robustness checks to ensure that our findings are causal in nature, i.e., they reflect the actual consequences of quarantining and do not simply mirror ongoing temporal or Reddit-wide trends.
Although we used a comprehensive set of lexicons to analyze trends of misogyny and racism, our analyses are limited by the selection of terms in these lexicons. It is likely that these communities have devised new misogynistic and racist terms that are not captured in the lexicons we use. How language evolves within these groups and how it shapes anti-feminist and racist discourse offers an interesting direction for future work.

\subsubsection{New v/s inactive users.} We reserved a buffer of 30 days before the start date of our analyses to identify new users. However, some users who we identify as \textit{new users} may not be new but just recently inactive. We computed that the average gap between two posts for any user has a median of 8 days. Thus, we expect that our buffer period of 30 days would ensure that most \textit{new users} are truly new. Even when that is not the case, the new engagement by previously inactive users speaks to the same issues as our analyses for new users, i.e., the extent to which quarantining affects the attraction of non-active members to the community.

\subsubsection{Facilitating discussion on race and masculinity.}
We note that we did not make the decision to quarantine TRP---Reddit administrators did. While our findings suggest their decision was wise, it is important to recognize that there are many topics about men and masculinity (e.g., high male suicide rates, society's neglect of male mental health) that need discussion in our society today.
Many scholars have emphasized the importance of recognizing multiple masculinities---black, white, working-class, middle-class---and the relations between them for an intersectional analysis of gender relations, reducing men's perpetration of violence, and increasing support for gender justice \cite{collins2004black,connell2005masculinities,courtenay2000constructions,peretz2016study}.
A crucial future direction is identifying how to support constructive discussions about masculinity and protect them from incursions of hate speech.
Similarly, the quarantine of TD helped reduce the influx of new users to a toxic community, and we consider this a desirable outcome. Still, we need to find ways to support important discussions about race and conservative principles that do not devolve into conspiracies, faux outrage over free speech, and racism.

\subsubsection{Potential for quarantining to increase insularity.}
Prior research has shown that mediated communication often facilitates the formation of exclusionary communities that in more extreme cases may become echo chambers \cite{pariser2011filter} where users intensify shared views, foster intolerance, and are sheltered from ideological opposition \cite{brainard2009cyber,garrett2009echo,jamieson2008echo,allison2020communal}.
By alienating sanctioned communities from the rest of the platform, quarantining may increase the insularity of those communities, reduce the wider public pressure to limit their regressive social attitudes, and encourage moves toward extremism. 
Isolated communities are also more likely to ``favor exit over voice \cite{centivany2016popcorn},'' i.e., they would rather exit the platform than express constructive criticism to help improve platform operations \cite{matias2016goingdark}.
Future research should therefore empirically evaluate the dynamics of how increased insularity within a community influences its members.

\subsubsection{Supporting community-wide moderation interventions.}
Finally, platforms like Reddit have difficult decisions to make when they choose to quarantine (or ban) a particular group.  While a European approach makes hate speech illegal, Americans tend to hesitate because choosing where to draw the line between offensive and not offensive speech can lead to a slippery slope.  Better data analysis and visualization tools could help administrators make more informed decisions about where and how to draw that line. 

It is also important to recognize that platforms do not take community-level moderation decisions unilaterally. These decisions are often a response to a long, drawn-out set of complaints by end-users and moderators, media rebukes, and pressure from the public, lawmakers and advertisers \cite{Carlson2020}.
Further, administrations' decisions to regulate controversial communities have previously resulted in widespread collective protests by moderators \cite{matias2016goingdark}---such incidents may influence how admins wield community-level actions in the future. 
Delineating the participatory aspect \cite{centivany2016popcorn} of such policy-making processes is an exciting direction for future research.

%% file: 06-implications.tex
\subsection{Design and Theoretical Implications}
\subsubsection{Effectiveness of design friction as a moderation strategy.}
For HCI, our work demonstrates the efficacy of introducing design friction \cite{Cox2016,Mejtoft2019} to counteract antisocial behavior in online communities.
Quarantining resulted in the following design changes: (1) TRP and TD were relegated to a ``sort of bad-actor closet \cite{cousineau2020displaced}'' through removal from aggregated content for non-community users; and (2) all visitors were shown a splash page saying that the community was quarantined and highly offensive, and asked if they would still like to view the community. This meant that participants had to deliberately choose to continue, in order to participate in the quarantined subreddit. 
While our methods do not allow us to disentangle the effects of these two changes, overall our findings show the potential for design friction to serve as a new moderation strategy in the toolbox of social platforms.
Given the philosophical debates around freedom of speech online \cite{Carlson2020}, and simultaneously, a platform's responsibility to protect its users and brand from harm \cite{ullmann2020quarantining}, quarantining provides a \textit{softer}, perhaps more ethical, alternative to banning.
Further, since community-wide moderation interventions like quarantining are necessarily human-made, they bypass the criticisms that platforms are increasingly facing about the pitfalls of automating their moderation \cite{Gillespie2020,Gorwa2020}.

\subsubsection{Framework to evaluate the effectiveness of interventions.}
For online moderation, we provide a computational framework that internet platforms can use to evaluate the effectiveness of moderation interventions.
{We would like to stress that our methods and causal framework are independent of the target community, and our code is publicly available.\footnote{We will make our code available after peer review.} This makes it easy to extend our approach to examine the effects of such interventions on other communities in the future. }

The empirical work in this paper suggests that when narrowly applied to specific, large misogynistic and racist groups, quarantining can work to reduce the influx of new users and contain the behavior.
Lately, governance scholars have used Reddit quarantining as a lever to discuss the ethics of content moderation by private companies \cite{Carlson2020,cousineau2020displaced}. For example, Carlson and Cousineau have explored existing ethical frameworks for content moderation such as the Santa Clara Principles \cite{santaclaraprinciples} and Rawls' \cite{rawls1971theory} veil of ignorance to develop a decision-making model that provides platforms guideposts for addressing complex content regulation questions.
We argue that the efficacy of such strategies, as demonstrated in this paper for quarantining, should also inform these ongoing conversations \cite{Carlson2020,cousineau2020displaced} around their possible future use.
Further, our framework can be easily adapted to evaluate other moderation interventions in the future.

\subsubsection{Platform interventions as graduated sanctions.}
In this work, we studied quarantining which is a community-level intervention. However, social platforms predominantly tend to use user-level interventions such as post removals (automated and manual), temporary banning, and suspensions. This raises the question: when should platforms rely on user-level interventions, and when should they resort to community-level interventions? 
In the distributed moderation design of Reddit, the local volunteer moderators of a community are largely responsible for implementing user-level interventions and the small group of paid Reddit administrators carry out community-level interventions~\cite{chandrasekharan2018norms}. This arrangement naturally leads to a hierarchy of graduated sanctions in which communities are punished increasingly severely---from removals of individual user postings to community bans. In their influential work ``Regulating behavior in online communities,'' Kiesler et al. theorized that ``graduated sanctions increase the legitimacy and thus the effectiveness of sanctions'' \cite{kiesler2012,ostrom1990governing}. In view of this, it is possible that community-level interventions will be more acceptable if they follow a stream of user-level interventions or corrections against local governance failures. 
We expect that even those communities that refuse to sanction users who violate Reddit's content standards may seek self-preservation and consider improving their moderation practices in the face of graduated sanctions such as quarantining.
Graduated sanctions can also help keep platforms accountable, ensuring that the reason for each escalation of sanctions is clear to users.
Indeed, in line with our findings, Reddit noted that the TD community did not reform their behavior, and consequently banned TD on June 29, 2020.

It is also important to note that quarantining is just one of the many levers of change available to platforms. Though actions like quarantining and banning are often widely covered by the news media \cite{peck2017,stewart_2019} and reflect well on the platform's moderation, they need to work alongside efficient user-level interventions and coordination with local community moderators to ensure overall platform health.

%% file: 07-conclusion.tex
\section{Conclusion}
In this paper, we examined the effectiveness of quarantining, a community-wide moderation intervention, in changing end-user behavior in two influential communities, the TRP and TD subreddits.
This low-cost intervention dramatically diminished the influx of newcomers to each community but it did not modulate the offensive discourse.
We conclude that quarantining can be a surprisingly effective compromise between censoring offensive speech and preserving freedom of expression in regulating content that exists on the margins of acceptability. 
Going forward, additional research and thoughtful design work can help to develop more moderation strategies that can achieve such compromises.